\documentclass[iop,english,twocolappendix,numberedappendix,showpacs,superscriptaddress,appendixfloats,tighten,apj,twocolumn]{aastex6}
\usepackage{amsmath,amsfonts,amssymb,xcolor,soul}
\usepackage{dcolumn,ulem}
\usepackage{bm}
\usepackage{float}
\usepackage{hyperref}
\usepackage{subfigure}

\usepackage{amstext}
\usepackage{graphicx}
\usepackage{color}

\def\bb{{\bf B}}
\def\vv{{\bf v}}

\def\be{\begin{equation}}
\def\ee{\end{equation}}
\def\ba{\begin{eqnarray}}
\def\ea{\end{eqnarray}}

\def \pmbtext#1{\leavevmode
     \setbox0\hbox{#1}
     \kern0,4pt \copy0 \kern-\wd0
     \kern-0,2pt \raise0,3pt \box0 }

\renewcommand{\baselinestretch}{1.2}
\newcommand{\el}{\boldsymbol{\ell}}
\newcommand{\va}{\textbf{b}}

\newcommand{\vh}{\textbf{v}}

\newcommand{\ja}{\textbf{j}}

\newcommand{\lang}{\left\langle}
\newcommand{\rang}{\right\rangle}

\newcommand{\nab}{\boldsymbol\nabla}
\newcommand{\dnu}{\textbf{d}_{\nu}}
\newcommand{\deta}{\textbf{d}_{\eta}}

\usepackage{soul}

\begin{document} 

\title{Fluid energy cascade rate and kinetic damping: new insight from 3D Landau-fluid simulations}

\author{R. Ferrand\altaffilmark{1}, F. Sahraoui\altaffilmark{1}, D. Laveder\altaffilmark{2}, T. Passot\altaffilmark{2}, P.L. Sulem\altaffilmark{2} and S. Galtier\altaffilmark{1,3}} 
\email{renaud.ferrand@lpp.polytechnique.fr}
\affil{$^1$ Laboratoire de Physique des Plasmas, CNRS, \'Ecole polytechnique, Universit\'e Paris-Saclay, Sorbonne Universit\'e, Observatoire de Paris-Meudon, F-91128 Palaiseau Cedex, France}
\affil{$^2$ Universit\'e C\^ote d'Azur, Observatoire de la C\^ote d'Azur, CNRS, Laboratoire J.L. Lagrange, Boulevard de l'Observatoire, CS  34229, 06304 Nice Cedex 4, France}
\affil{$^3$ Institut Universitaire de France (IUF)}

\date{\today}


\begin{abstract}
Using an exact law  for incompressible Hall magnetohydrodynamics (HMHD) turbulence, the energy cascade rate is computed from three-dimensional HMHD-CGL (bi-adiabatic ions and isothermal electrons) and Landau fluid (LF) numerical simulations that feature different intensities of Landau damping over a broad range of wavenumbers, typically $0.05\lesssim k_\perp d_i \lesssim100$. Using three sets of cross-scale simulations where turbulence is initiated at large, medium and small scales, the ability of the fluid energy cascade to ``sense'' the kinetic Landau damping at different scales is tested. The cascade rate estimated from the exact law and the dissipation calculated directly from the simulation are shown to reflect the role of Landau damping in dissipating energy at all scales, with an emphasis on the kinetic ones. This result provides new prospects on using exact laws for simplified fluid models to analyze dissipation in kinetic simulations and spacecraft observations, and new insights into theoretical description of collisionless magnetized plasmas. 
\end{abstract}

\maketitle

\section{Introduction}

The understanding of turbulent astrophysical plasmas remains to date a challenging problem: their chaotic nature and the complexity of the mechanisms at work in such media impose limitations to the methods one can use to study them efficiently. Yet, enhancing our understanding of turbulent plasmas would provide the keys to solve a variety of problems related to energy dissipation, particle heating and acceleration. Examples of systems where these processes are crucial include the solar wind (SW) and planetary magnetospheres \citep{Bruno05,Matthaeus11,Goldstein95,Sahraoui20}, accretion flows around compact objects \citep{Balbus98,Quataert99} and fusion devices \citep{Diamond05,Garbet06,Fujisawa21}. In the SW, the heating problem is reflected by the slow decline of ion temperature (as function of the radial distance from the sun) in comparison with the prediction from the adiabatic expansion model of the wind \citep{Richardson95}. Turbulence has long been proposed as a way to explain this behavior \citep{Matthaeus99}, through scale-by-scale transfer of energy (i.e., cascade) toward small (kinetic) scales where dissipation is more effective \citep{Schekochihin09}. A common tool used to estimate this energy dissipation is the formalism of exact law for fully developed turbulence first introduced by \citet{kolmogorov} to study incompressible neutral fluids. In this formalism, energy is assumed to be injected at large scales at a constant rate per unit volume $\varepsilon$, which is assumed to be equal to the rate of cascade to smaller scales and to the rate of dissipation at those scales. Assuming statistical homogeneity and stationarity of the turbulent fields, and the existence of an inertial range in which both forcing and dissipation mechanisms are negligible, the cascade rate $\varepsilon$ must remain constant in the inertial range \citep{kolmogorov,monin57,antonia97}. The formalism of exact law has been extended to (in)compressible magnetized plasmas within various approximations
\citep{PP98,galtier08,Banerjee13,A2017b,hellinger18,andres18,ferrand19}.

Exact laws have been used successfully to measure the energy cascade rate in the SW~ \citep{Smith06,Podesta07,sorriso07,macbride08,Marino08,Carbone09,smith09,Stawarz,Osman,coburn,BanerjeeSW16,hadid17} and terrestrial magnetosheath \citep{hadid18,andres19}. In those studies, the estimated cascade rate was interpreted as the turbulence energy  dissipation rate, and hence used to quantify the amount of plasma heating due to turbulence \citep{sorriso07,Carbone09,BanerjeeSW16}. However, as explained above, such an equivalence between injection, cascade and dissipation rates stems only from the hypothesis underlying exact laws derivation and cannot be demonstrated in spacecraft observations. Indeed, while in numerical simulations the injection, cascade and dissipation rates can generally be estimated separately and compared to each other as done in this paper, estimating ({\it irreversible}) dissipation from spacescraft observation is a challenge and, generally, only the cascade rate, which is directly linked to measurable quantities through the exact law, is accessible \citep{sorriso07,hadid17}. Thus, in spacecraft data, interpreting the energy cascade rate as the actual dissipation rate is not straightforward. This is particularly true because of the weakly collisional nature of the SW: in such plasmas classical viscous and/or resistive effects are absent, and dissipation is expected to occur via kinetic effects (e.g., Landau and cyclotron resonances) \citep{leamon98,Sahraoui09,sahraoui10,He15,Chen19} that are not captured by usual fluid descriptions of plasmas. A fundamental question arises here: is the {\it fluid} turbulent cascade rate estimated in simulations and spacecraft observations of space plasmas representative of the actual {\it kinetic} dissipation in those media? It is the main goal of this paper to address this question, which impacts the use of fluid models to interpret part of in-situ spacecraft observations in the near-Earth space and the theoretical (fluid vs. kinetic) modeling of weakly collisional plasmas. 
In contrast with previous studies based on 2D hybrid particle-in-cell simulations \citep{hellinger18, Bandyopadhyay20}, the use of 3D LF models  give the possibility to isolate the influence of electron and ion Landau damping, neglecting all the other kinetic effects, and is therefore very suited to address the question of interest here.

\section{Theoretical model} \label{theory}
Although we are dealing with weakly compressible regimes we chose, for the sake of simplicity, to use here the exact law derived by \citet{ferrand19} for incompressible HMHD (see below about the use of more general compressible models). Starting from the incompressible HMHD equations, and under the usual assumptions of time stationarity, space homogeneity and infinite (kinetic and magnetic) Reynolds numbers,  one can derive for the energy cascade rate in the inertial range the expression $\varepsilon = \varepsilon^{MHD} + \varepsilon^{Hall}$, with
\begin{align} \label{IMHD}
	\varepsilon^{MHD} =& -\frac{1}{4}\nab_{\el}\cdot \lang (|\delta\vh|^2 + |\delta\va|^2)\delta\vh -2(\delta\vh\cdot\delta\va)\delta\va \rang, \\ \label{Hall}
	\varepsilon^{Hall} =& -\frac{1}{8}d_i\nab_{\el}\cdot \lang 2(\delta\va\cdot\delta\ja)\delta\va - |\delta\va|^2\delta\ja \rang,
\end{align}
where $\vh$, $\va = \bb / \sqrt{\mu_0 \rho_0}$ and $\ja = \nab \times \va$ are the velocity,  magnetic field and electric current in Alfv\'en units ($\rho_0$ is the constant mass density) and $d_i$ is the ion inertial length. Fields are taken at points $\textbf{x}$ and $\textbf{x}'$ separated by a spatial increment
$\el=\textbf{x}' -\textbf{x}$, and the notations $\vv \equiv \vv(\textbf{x})$ and $\vv' \equiv \vv(\textbf{x}')$ are adopted. We then define the increment operator $\delta$ as $\delta \vv = \vv' - \vv$, and $\nab_{\el}$ as the derivative operator with respect to the increment $\el$. 

\section{Simulation data} \label{simulation}

\subsection{Presentation of the data}

In this study, HMHD-CGL refers to a fluid model with anisotropic ion pressure whose gyrotropic components parallel and perpendicular to the local magnetic field obey nonlinear dynamical equations where the heat fluxes are neglected (bi-adiabatic approximation introduced by Chew, Goldenberg and Low \citep{CGL56}, thus the acronym). The electrons are assumed  isothermal. Differently, the LF model retains the nonlinear dynamics of the parallel and perpendicular pressures and heat fluxes for both the ions and electrons, and involves a closure at the level  of the fourth-order moments, consistent with the low-frequency linear kinetic theory \citep{SHD97,Passot07}. The main assumption for modeling Landau damping consists in retaining the imaginary contribution of the plasma response function in the closure relation which expresses the last retained fluid moment of the hierarchy in terms of  the lower ones. In Fourier space, this procedure generates factors of the form ``$i \,{\rm sgn} (k_z)$" which, in physical space, identifies with the Hilbert transform along the ambient magnetic field \citep{hammett90,hunana19b}. It is then possible to  generalize this formulation to take into account magnetic field line distortion, using the convolution form of the Hilbert transform \citep{SHD97}. Its approximation in the numerical code is discussed in \citet{Passot14}. In both models, finite ion and electron Larmor radius corrections are neglected, thus reducing the kinetic effects to Landau damping. The Ohm's law includes the Hall term and the electron pressure contribution. Turbulence is forced with counter-propagating kinetic Alfv\'en waves (KAWs) making an angle $\theta$ with the ambient magnetic field, at the largest scales of the simulation domain. This corresponds to transverse wavenumbers $k_{\perp,f}$, whose values are summarized in Table \ref{runs}. The amplitudes obey a Langevin equation, with an oscillation frequency given by the KAW linear dispersion relation \citep{TENBARGE14}. We also introduce two thresholds in order to constrain the sum of perpendicular kinetic and magnetic energies to stay within a certain range. Small-scale dissipation is ensured by the hyperviscosity and hyperdiffusivity terms in the velocity and induction equations, of the form $\dnu = \nu(\Delta_\perp +\alpha \partial_z^2)^4 \vv$ and $\deta = \eta(\Delta_\perp +\alpha \partial_z^2)^4 \va$, with $\alpha$ being an anisotropy coefficient.

In all the simulations, $\beta_i = 1$ and the ion and electron pressures are taken isotropic and equal initially. The other parameters are reported in Table \ref{runs}. The simulations are performed using a desaliased spectral code (at 2/3 of the maximum wavenumber) with a third-order Runge-Kutta scheme for time stepping.

\begin{table}
\centering
\begin{tabular}{ m{3em} m{3em} m{5em} m{1.5em} m{5em} m{2em} } 
 \hline
 Run & $k_{\perp,f}d_i$ & Resolution & $\theta$ & $\nu=\eta$ & $\alpha$ \\ 
 \hline
 CGL1 & $0.045$ & $512^3$			  & $83^{\circ}$ & $7.35\times10^{-8}$ 	& 80\\ 
 CGL2 & $0.045$ & $512^3$		      & $75^{\circ}$ & $7.35\times10^{-8}$ 	& 10\\ 
 CGL3 & $0.5$  & $512^2\times1024$  & $75^{\circ}$ & $10^{-14}$ 			& 2.5\\ 
 CGL4 & $0.011$ & $1024^3$           & $75^{\circ}$ & $3\times10^{-3}$		& 5\\ 
 LF1  & $0.045$ & $512^3$            & $83^{\circ}$ & $7.35\times10^{-8}$	& 1\\ 
 LF2  & $0.045$ & $512^3$            & $75^{\circ}$ & $7.35\times10^{-8}$	& 1\\ 
 LF3  & $0.5$  & $432^3$            & $75^{\circ}$ & $7\times10^{-14}$		& 1.5\\ 
 LF4  & $0.011$ & $512^3$            & $75^{\circ}$ & $3\times10^{-3}$		& 2\\ 
 \hline
\end{tabular}
 \caption{List of runs and their relevant parameters, where CGLx and LFx refer to HMHD-CGL and LF simulations, respectively. The ratio of the longitudinal to transverse box sizes is given by $\tan(\theta)$.}
\label{runs}
\end{table}

\begin{figure}
\centering
\includegraphics[width=\hsize,trim={30 0 110 35},clip]{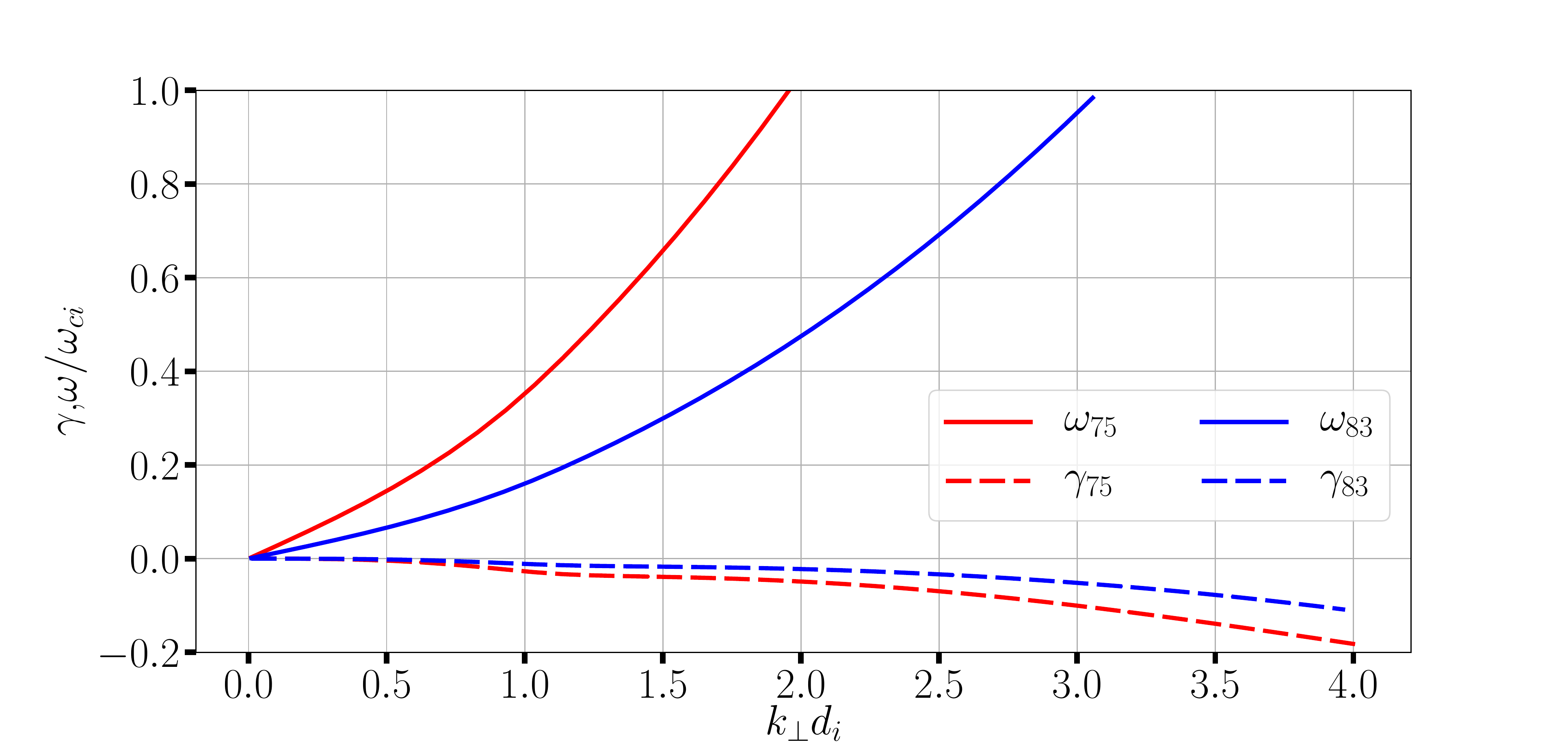}
\caption{Frequency $\omega$ and damping rate $\gamma$ (normalized by the ion gyrofrequency $\omega_{ci}$) of KAWs versus the normalized transverse wavenumber $k_\perp d_i$ (where $d_i$ refers to the ion inertial length) for the LF model at the two  propagation angles $\theta=83^\circ$ and $75^\circ$ used in driving the simulations ($\beta_i=1$, $T_i=T_e$).}
\label{disp-whamp}
\end{figure}

Two different propagation angles for the KAWs driven at the largest scale of each LF simulation were chosen, hence tuning Landau damping to two different levels \citep{kobayashi17}. This can be seen in Fig.~\ref{disp-whamp} which compares the linear dispersion relation and damping rate of the KAWs: the higher the propagation angle, the lower the damping rate at a given scale. Note that, while changing the angle, we do not change the amplitude of the fluctuations at the driving scale (i.e., $\omega_{NL}$ remains constant), and so the nonlinear parameter $\chi= \omega_{NL}/\omega_{L}$ also varies: when the angle decreases, $k_{\parallel}$ increases and so does $\omega_{L}$, thus $\chi$ is reduced. As the ratio $\gamma/\omega_{L}$ is approximately constant for high oblique angles $\theta$ (e.g., Fig. 7 in \citet{sahraoui12}), the strength of the Landau damping relative to the cascade rate $\gamma/\omega_{NL} = (1/\chi) (\gamma/\omega_{L})$  thus increases as the angle decreases.

\subsection{Energy balance and time stationarity}

Let ${\mathcal E}_{{\rm tot}}(t)$ be the total energy of the system at time $t$, $ {\mathcal I}_{\rm t}(t)$ the injection rate due to the external forcing on the perpendicular velocity components, and ${\mathcal D}_{\rm h}(t)$ the total dissipation rate due to the hyperviscous and hyperdiffusive terms.
Since Landau damping does not affect the total energy balance, total energy conservation implies
\begin{equation}
\frac{d}{dt} {\mathcal E}_{{\rm tot}}(t) = {\mathcal I}_{\rm t}(t) -{\mathcal D}_{\rm h}(t).
\label{eq:totalenergy}
\end{equation}
Denoting by ${\mathcal E}_{{\rm int}}$ and ${\mathcal E}_\|$ the parts of the total energy associated with the pressure components (internal energy) and the parallel velocity and magnetic field components entering the kinetic and magnetic parts, we can write
\begin{equation}
\frac{d}{dt} {\mathcal E}_{{\rm tot}}(t) -\frac{d}{dt} {\mathcal E}_{{\rm int}}(t)- \frac{d}{dt} {\mathcal E}_\|(t)\equiv\frac{d}{dt} {\mathcal E}_\perp (t)\approx 0,
\end{equation}
where ${\mathcal E}_\perp $ is the sum of the perpendicular kinetic and magnetic energies, a quantity bound to remain nearly constant by the forcing procedure.

\begin{figure}
	\centering
	\includegraphics[width=\hsize,trim={50 10 10 25},clip]{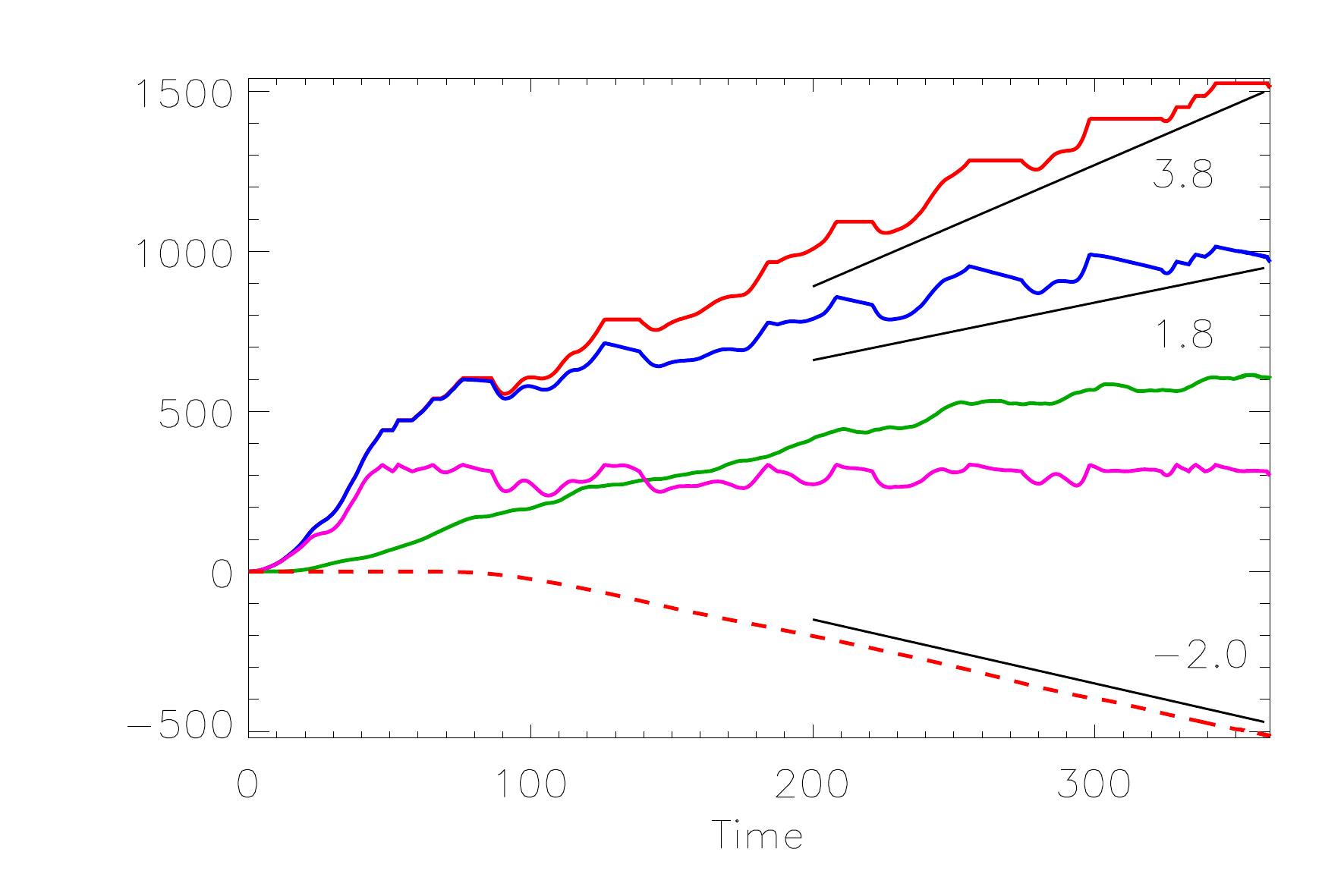}
	\includegraphics[width=\hsize,trim={50 10 10 25},clip]{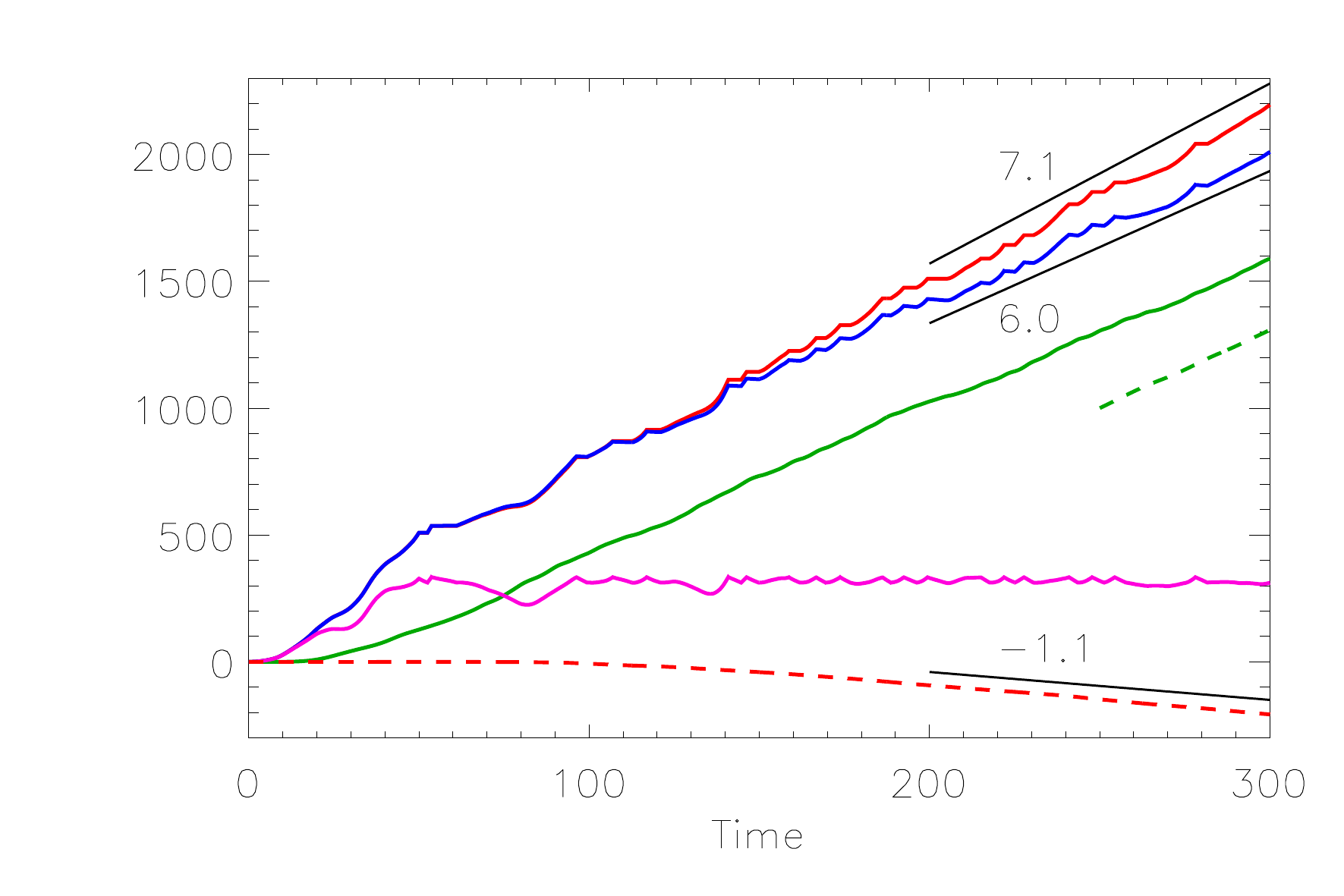}
	\caption{Low resolution runs of CGL3 (CGL3-LR, top) and LF3 (LF3-LR, bottom): Time evolution of the total energy injected in the system  $\int_0^t{\mathcal I}_{\rm t} dt$ (solid red line), total energy ${\mathcal E}_{{\rm tot}}$ (solid blue), internal energy ${\mathcal E}_{{\rm int}}$ (solid green), perpendicular energy ${\mathcal E}_{\perp}$ (solid magenta, roughly constant) and the time integrated hyperdissipation $\int_0^t{\mathcal D}_{\rm h} dt$ (dashed red). The piece of dashed green curve (right) starting at $t=250$, whose vertical position is arbitrary, displays the heating due to heat fluxes, which is consistent with the increase of internal energy.}
	\label{fig:lowres_lfcgl}
\end{figure}

Because of computational constraints, the time evolution of the different energy components is computed for low resolution (LR) simulations analogs of runs CGL3 and LF3 and shown in Fig. \ref{fig:lowres_lfcgl} (injection and dissipation rates needed to perform this extra study were not output at a high-enough frequency in the large-resolution simulations).
From this figure it is conspicuous that the time evolution of total energy, injection and hyperdissipation is consistent with the energy conservation (\ref{eq:totalenergy}). Moreover, one can see the driving procedure at play in keeping the perpendicular energy ${\mathcal E}_\perp$ roughly constant. Its time stationarity is in practice established when the hyperdissipation rate has reached a constant value. When comparing CGL3-LR with LF3-LR, one notices that run LF3-LR requires a larger injection rate to maintain the same level of turbulence on the magnetic and perpendicular velocity than in run CGL3-LR, since Landau damping efficiently converts a part of the injected energy into internal energy. This is evidenced by the dashed green curve in Fig. \ref{fig:lowres_lfcgl} (bottom), which shows that the increase of the internal energy is consistent with the heating by heat fluxes. Moreover, the hyperdissipation rate is lower on run LF3-LR, suggesting that part of the cascading energy is taken by Landau damping, as will be evidenced in next section.

\section{Calculation of the energy cascade rate} \label{results}

Using Eqs. (\ref{IMHD})-(\ref{Hall}), we compute the transverse energy cascade rate for each simulation as a function of the perpendicular increment by averaging over all spatial positions in the simulation box and different increment vectors, at a time for which the simulations reached a stationary state. Increment vectors $\el=(\ell_\perp,\ell_\parallel)$ are selected following the angle averaging method of \citet{taylor03}, and only the increments forming an angle of at least $45^{\circ}$ with the parallel direction are retained. As already mentioned, the exact law used is the one for incompressible HMHD, whereas the simulations are weakly compressible. A comparison (not shown) with a full compressible HMHD exact law \citep{andres18} only showed slight change ($\lesssim 10\%$ in the inertial range) of the cascade rate with respect to the current estimate from the incompressible model. The transverse cascade rate is then averaged over all increments of equal value of $\ell_\perp$.
%
%
The transverse hyperdissipation is computed in Fourier space as 
\begin{equation}
    \varepsilon^{diss} (\ell_\perp) = \int_0^{k_{\perp }}  dk'_\perp \int  k'^8_\perp (\eta |\textbf{b}({\boldsymbol k}')|^2 + \nu |\textbf{v}({\boldsymbol k}')|^2)  k'_\perp d\theta' dk'_z
\end{equation}
where we use $\ell_\perp = \pi/k_\perp$.

\begin{figure}
\centering
\includegraphics[width=\hsize,trim={65 65 110 70},clip]{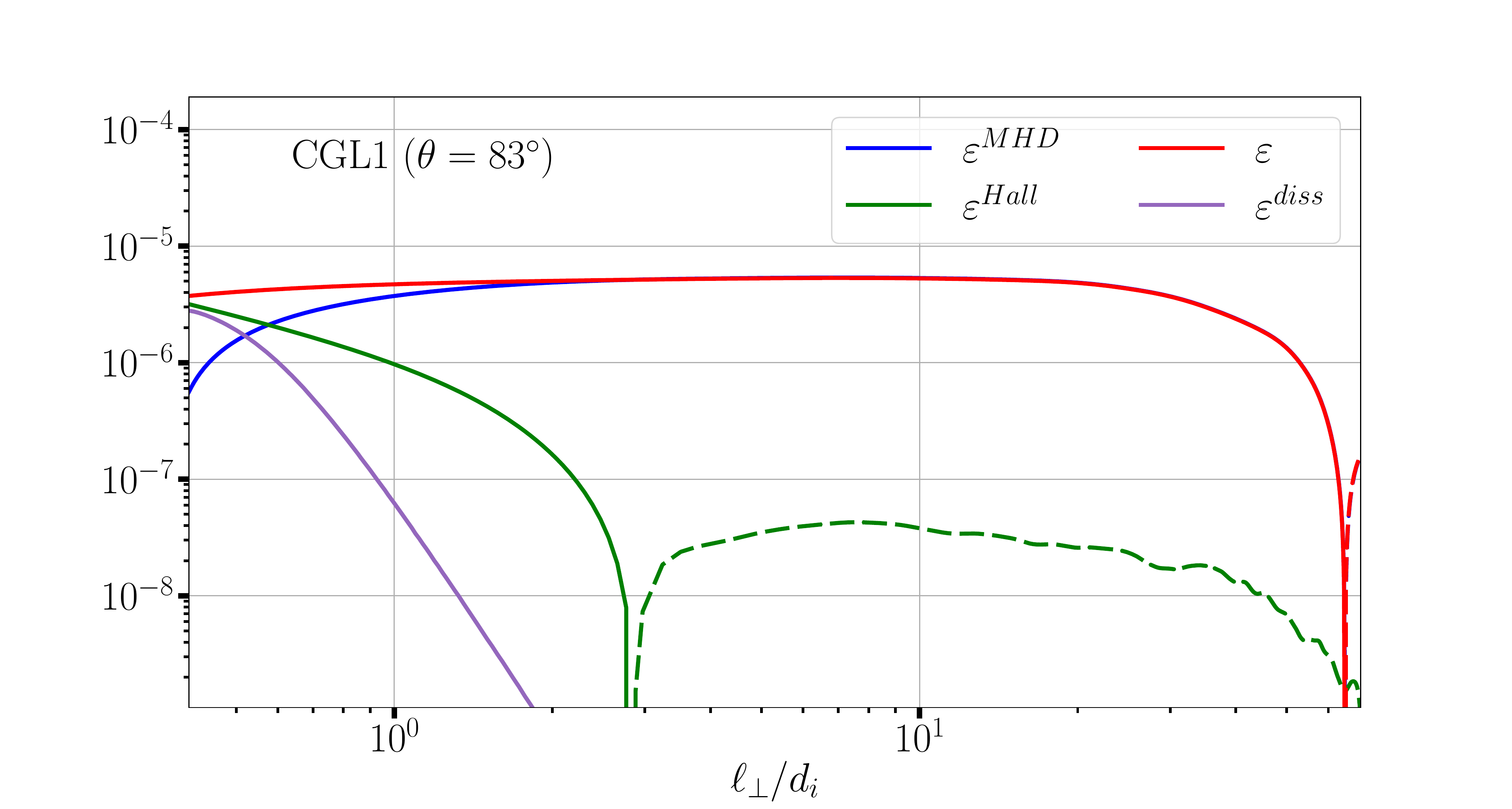}
\includegraphics[width=\hsize,trim={65 0 110 70},clip]{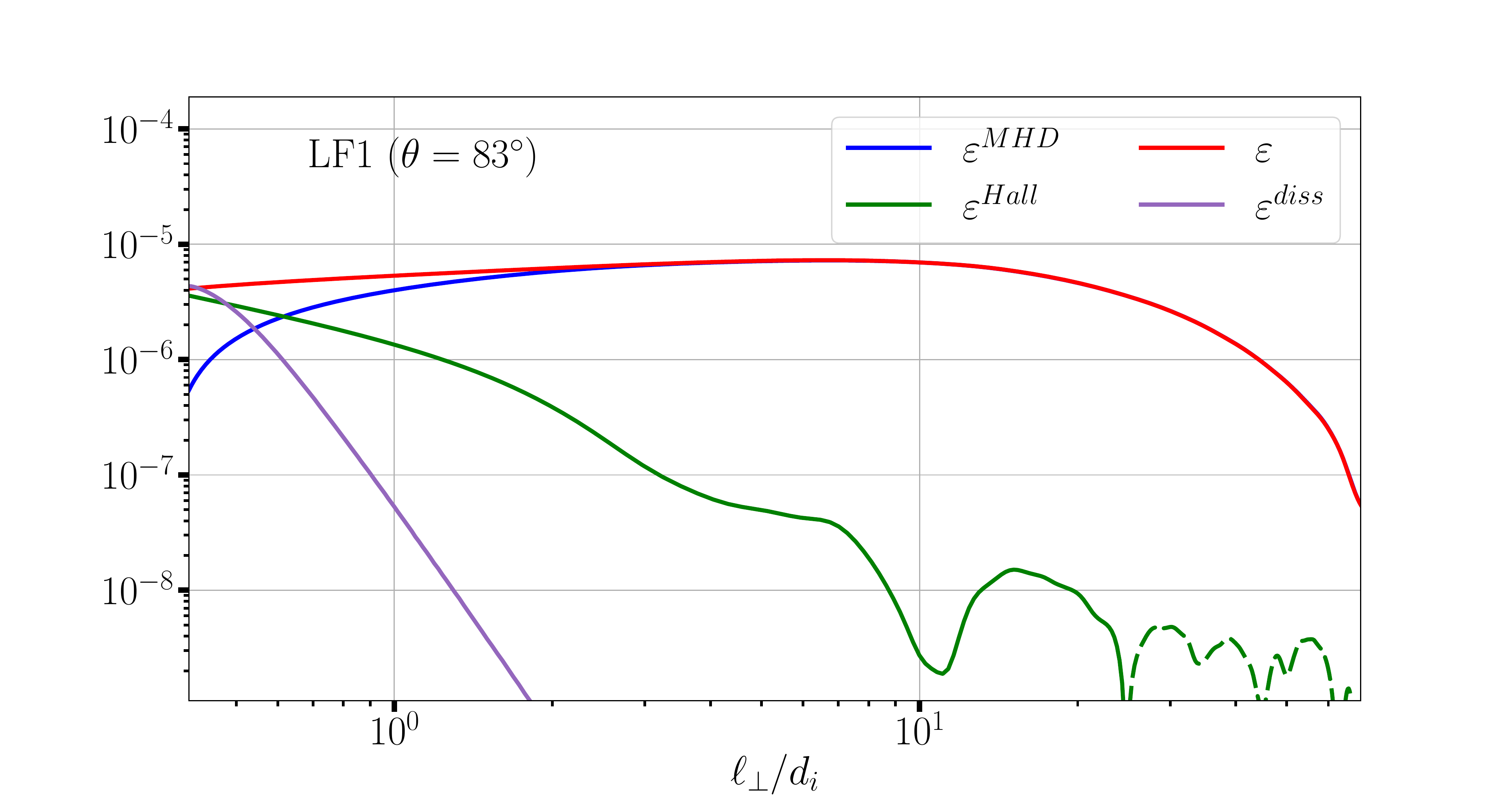}
\caption{Energy cascade rate $\varepsilon$, its ideal MHD and Hall components together with the transverse hyperdissipation computed for runs CGL1 (top) and LF1 (bottom). Plain lines represent positive values and dashed lines negative values.}
\label{med}
\end{figure}

\begin{figure}
\centering
\includegraphics[width=\hsize,trim={70 0 110 30},clip]{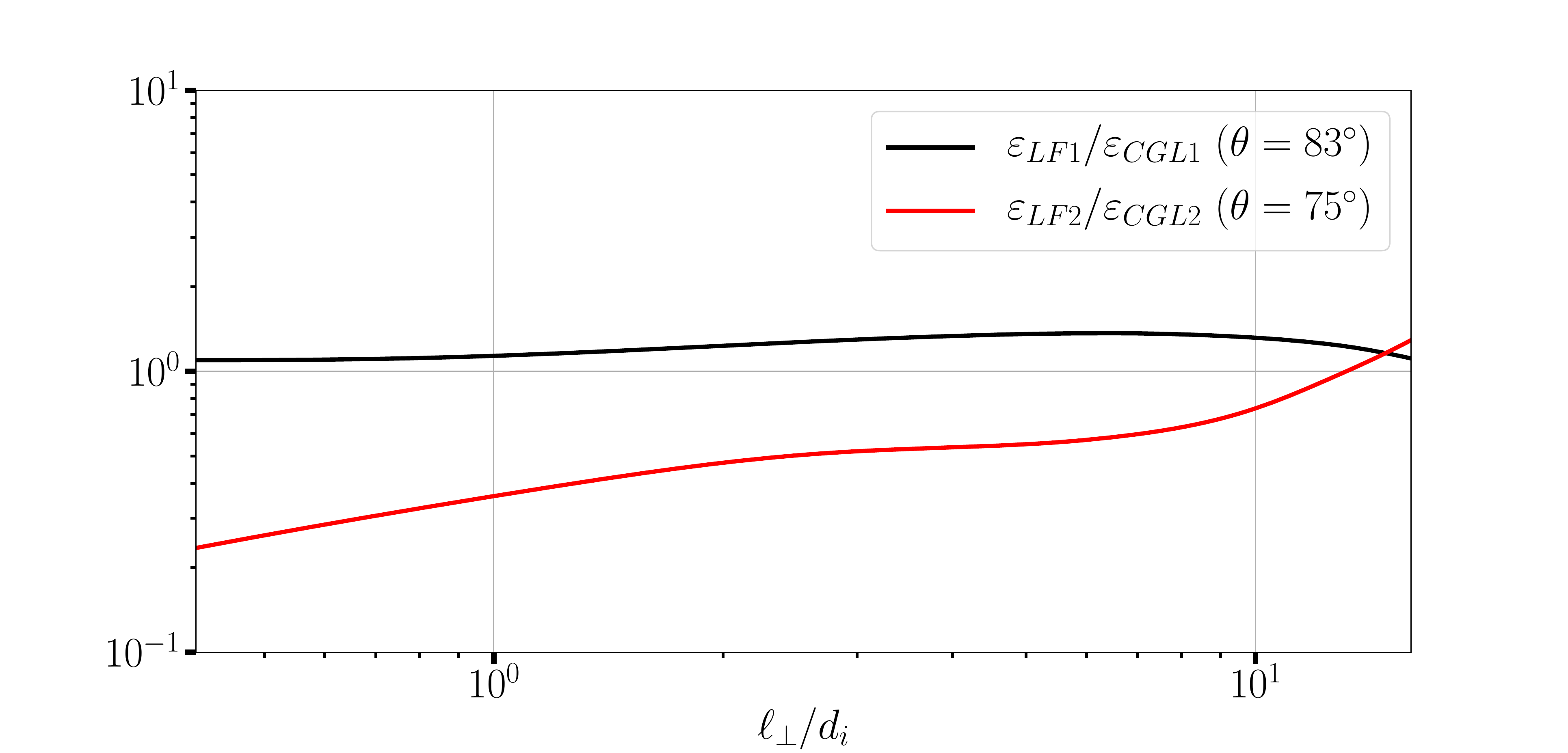}
\caption{Ratios of the energy cascade rate computed for LF simulations over the one for CGL simulations for a driving wave angle  $\theta=83^{\circ}$ (black) and $\theta=75^{\circ}$ (red).}
\label{ratio_med}
\end{figure}

For simulations forced at intermediate scales, whose results are reported in Fig. \ref{med}, the behavior of the MHD and Hall contributions to the energy cascade rate are similar, with the latter rising up at sub-ion scales, then dominating the former at about the ion inertial length. The total energy cascade rate is roughly constant on more than one decade of scales in the simulations with $\theta=83^{\circ}$, in particular in CGL1, which demonstrates the existence of an inertial range. To highlight the effect of Landau damping on the cascade rate, we compare in Fig. \ref{ratio_med} the cascade rates from the LF simulations with $\theta=75^{\circ}$ and $\theta=83^{\circ}$, normalized to the corresponding ones from the CGL simulations. We observe a stronger decrease (by up to a factor 5) in the normalized cascade rate at small scales for LF2 ($\theta = 75^{\circ}$), i.e. for the simulation with the strongest Landau damping, than for LF1 ($\theta = 83^{\circ}$) for which the normalized cascade rate remains nearly constant at all scales. This result clearly relates the enhancement of Landau damping at kinetic scales to the decline of the energy cascade rate at these scales. We note also the consistency between the (transverse) hyperdissipation and the cascade rate at the smallest scale of the simulation box (Fig. \ref{med}). 

\begin{figure}
\centering
\includegraphics[width=\hsize,trim={65 65 90 55},clip]{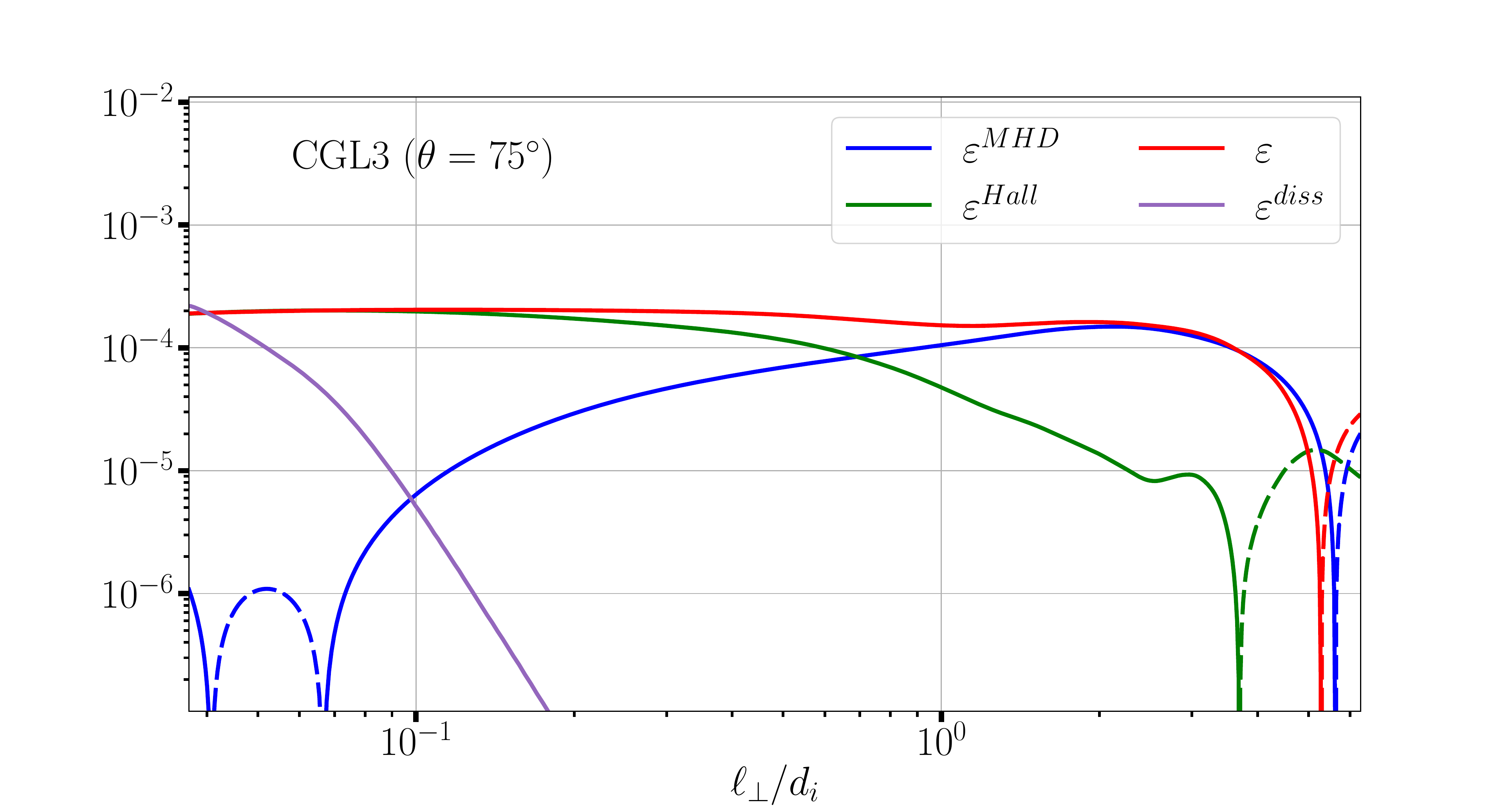}
\includegraphics[width=\hsize,trim={65 0 90 60},clip]{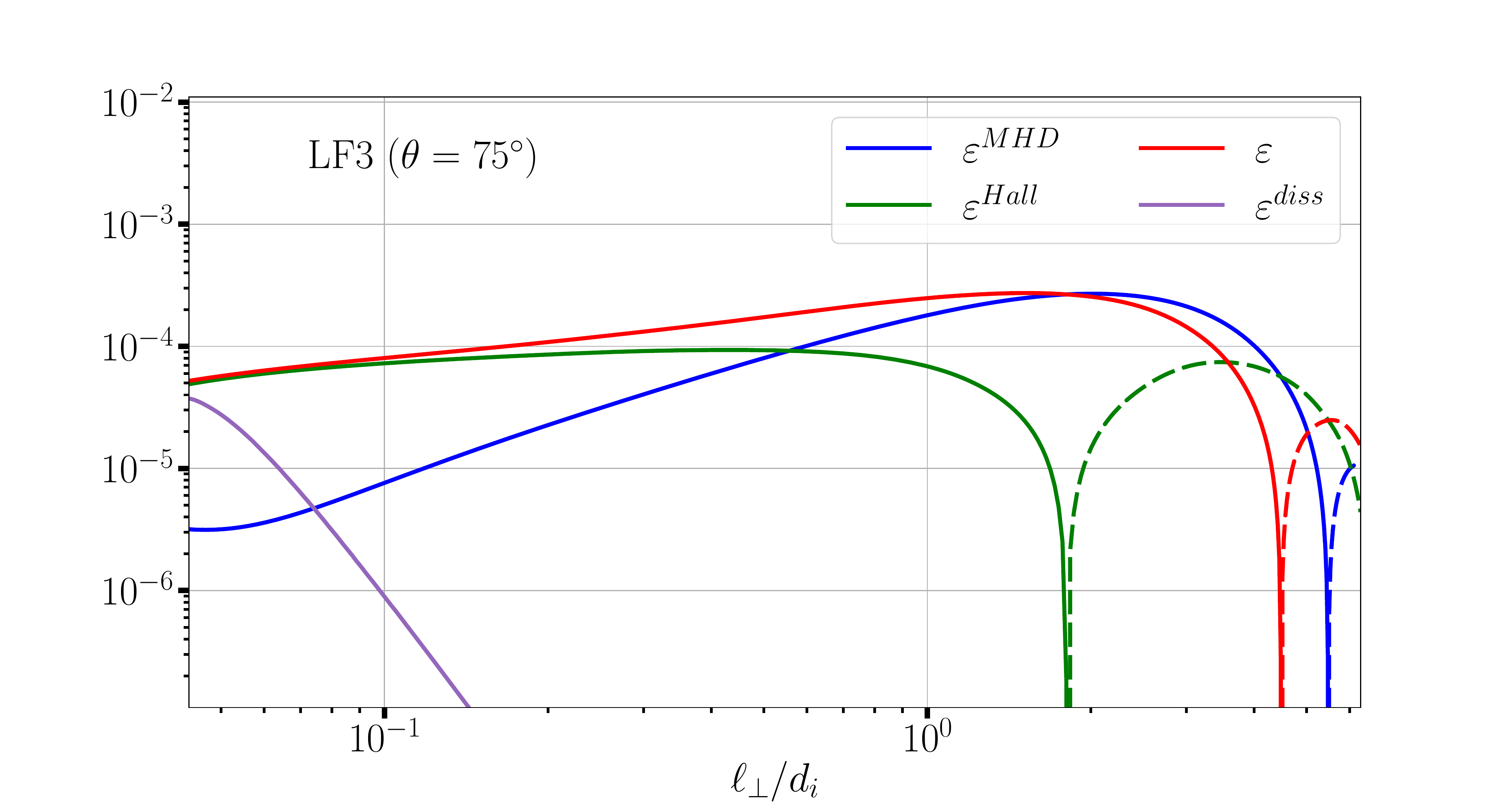}
\caption{Same as in Fig. \ref{med} for runs CGL3 (top) and LF3 (bottom).}
\label{small}
\end{figure}

We complement our study with the cascade rates estimated from simulations forced at even smaller scales (LF3 and CGL3) with $\theta=75^{\circ}$ and reported in Fig. \ref{small}. Simulation LF3 exhibits a strong decrease in $\varepsilon^{MHD}$ partially compensated by a quick rising of the Hall component, giving no clear inertial range, in contrast to CGL3 which still behaves similarly to the simulations forced at intermediate scales. As shown below, this effect may be attributed to the fact that Landau dissipation reaches high levels at the sub-ion scales of LF3, whereas CGL3 contains no dissipation mechanisms other than hyper viscosity and diffusivity, which are bound to act only at the smallest scales. Note that the sudden changes of sign observed at large scales in some components of the cascade rates in Figs. \ref{med} and \ref{small} are likely to be due to the proximity of the forcing. Those observed at small scales for the MHD component of run CGL3 would result from numerical errors in the calculation of $\varepsilon^{MHD}$ given its very small magnitude at those scales.

\begin{figure}
\centering
\includegraphics[width=\hsize,trim={70 0 110 30},clip]{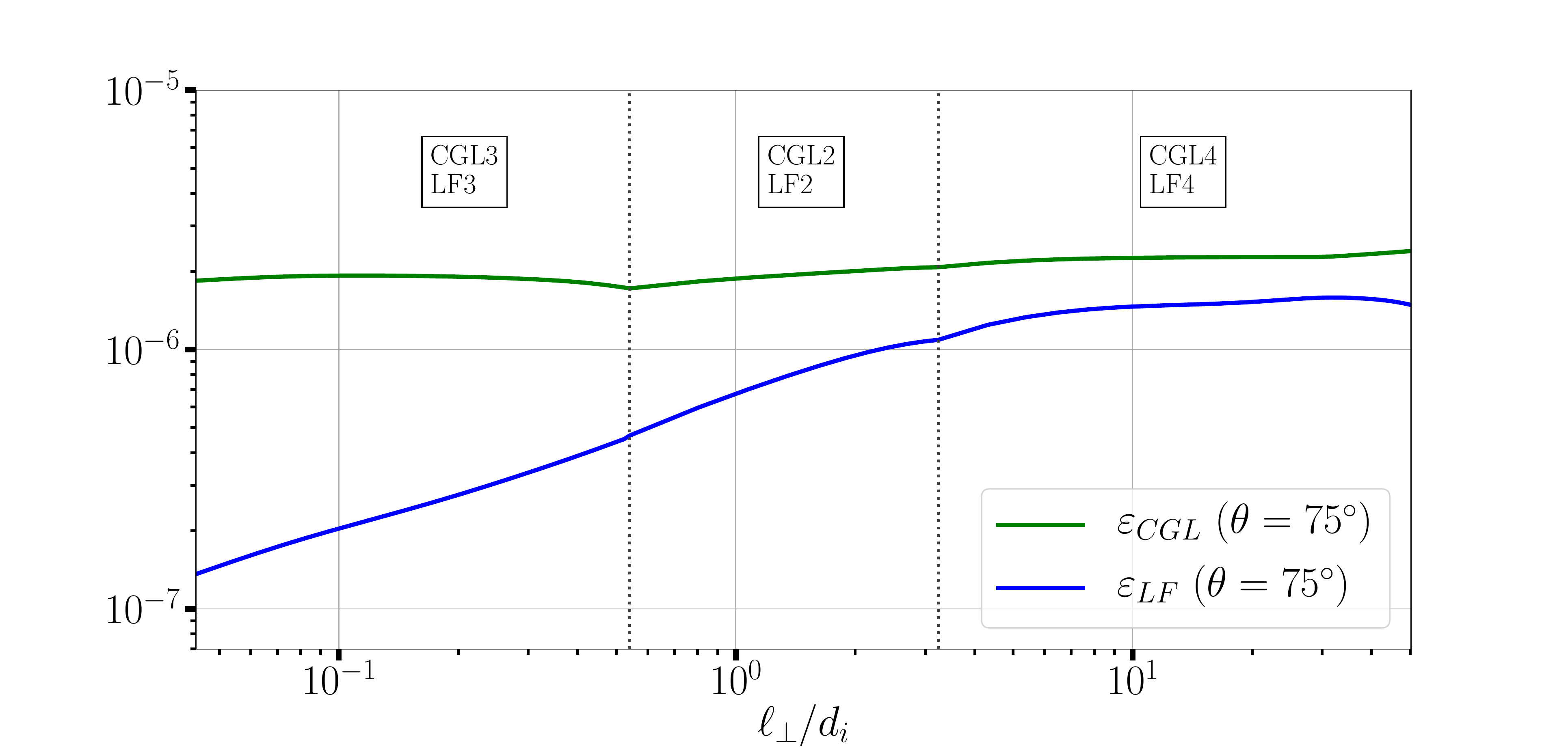}
\caption{Energy cascade rates reconstructed with CGL2-3-4 runs and LF2-3-4 runs. The ranges spanned by each simulation are delimited by the black dotted lines. A slight irregularity is observed on the green curve at the transition between CGL2 and CGL3, which is caused by an insufficient overlap of the cascade rates at these scales.}
\label{reconstructed}
\end{figure}

To obtain a full picture as to how Landau damping affects the energy cascade rate, we performed simulations forced at large scales (LF4 and CGL4). Combining the runs CGL2-3-4 and LF2-3-4 we construct a multi-scale energy cascade rate over nearly three decades of scales that highlights the effect of Landau damping on it.  As the simulations were run at different scales, the amplitude of the forcing was changed to ensure that each simulation reaches a fully turbulent state. Therefore, we renormalized the cascade rate $\varepsilon$ obtained from the different simulations to match the one of intermediate runs CGL2 and LF2, while taking care to discard the smallest scales of intermediate and large-scale forcing cascade rates to ensure that hyperviscosity is not acting at intermediate scales of the reconstructed energy cascade. Fig. \ref{reconstructed} shows the full energy cascade rate for CGL and LF runs for the driving wave angle $\theta=75^{\circ}$. CGL runs exhibit an almost constant energy cascade rate over two and a half decades of scales, whereas $\varepsilon_{LF}$ decreases steadily over scales and reaches its minimum value at the smallest ones, confirming that the behavior already observed in Fig. \ref{ratio_med} remains valid over a broader range of scales.

\section{Influence of Landau dissipation}

\subsection{Heating due to heat fluxes}

In the wake of the previous results an important question arises : can the drop in the energy cascade rate for LF runs be directly connected to Landau damping? For this purpose, we calculate the heating due to heat fluxes in presence of Landau damping. For each species, the pressure equations with the Hall term and the gyrotropic heat fluxes read

\begin{align}
\frac{d}{dt}\ln\left(\frac{p_\| |B|^2}{\rho^3}\right)&=-\frac{2c}{|B|}\widehat{\textbf{b}}\cdot\nab \times \boldsymbol{E_H} \nonumber \\
&-\frac{1}{p_\|} \left ( -2q_\perp \nab \cdot \widehat{\textbf{b}}+\nab\cdot (q_\| \widehat{\textbf{b}})\right), \label{eq:spar}\\
\frac{d}{dt}\ln\left(\frac{p_\perp }{\rho |B|}\right)&=\frac{c}{|B|}\widehat{\textbf{b}}\cdot\nab \times \boldsymbol{E_H} \nonumber \\
&-\frac{1}{p_\perp} \left ( q_\perp \nab \cdot \widehat{\textbf{b}}+\nab\cdot (q_\perp \widehat{\textbf{b}})\right ).\label{eq:sperp}
\end{align}

We define the parallel, perpendicular and total entropies per unit mass
\begin{equation}
s_\|=\frac{c_V}{3}\ln (\frac{p_\| |B|^2}{\rho^3}), ~ s_\perp=\frac{2c_V}{3}\ln(\frac{p_\perp }{\rho |B|}),
\end{equation}
\begin{equation}
s=s_\|+s_\perp=\frac{c_V}{3}\ln (\frac{p_\| p_\perp^2}{\rho^5}),
\end{equation}
where $c_V$ is the specific heat at constant volume.
Denoting by $e$ the internal energy per unit mass, the internal energy per unit volume reads $E\equiv\rho e=p_\perp + \frac{1}{2}p_\|=\frac{3}{2}nT$ where $T=\frac{1}{3}(2T_\perp + T_\|)$. From $e=c_V T$, one gets $c_V=\frac{3}{2m}$ (the Boltzmann constant is included in the definition of temperature). The total entropy then obeys
\begin{widetext}
\begin{equation}
\partial_t (\rho s) +\nab\cdot \left (\rho s \boldsymbol{u}+(\frac{q_\perp}{T_\perp}+\frac{q_\|}{2T_\|})\widehat{\textbf{b}}\right )=(\frac{1}{T_\|}-\frac{1}{T_\perp})q_\perp \nab \cdot \widehat{\textbf{b}} -\left ( \frac{q_\perp}{T_\perp} (\widehat{\textbf{b}}\cdot\nab)\ln T_\perp +\frac{q_\|}{2T_\|}(\widehat{\textbf{b}}\cdot\nab)\ln T_\| \right ). 
\end{equation}
\end{widetext}

From the form of the right hand side of Eqs. (\ref{eq:spar})-(\ref{eq:sperp}), we can conclude that the rates of change of the parallel and perpendicular entropies  per unit mass ($s^p_\|$ and $s^p_\perp$ respectively) associated with a production (or destruction) and excluding transport or exchanges between the parallel and perpendicular directions (see e.g. \citet{Hazeltine2013}),  are given by 
\begin{align}
&	\frac{d}{dt}s_\|^p=\frac{1}{\rho T_\|}q_\perp \nab \cdot \widehat{\textbf{b}}-\frac{q_\|}{2\rho T_\|}(\widehat{\textbf{b}}\cdot\nab)\ln T_\| \\
&	\frac{d}{dt}s_\perp^p=-\frac{1}{\rho T_\perp}q_\perp \nab \cdot \widehat{\textbf{b}}- \frac{q_\perp}{\rho T_\perp} (\widehat{\textbf{b}}\cdot\nab)\ln T_\perp. 
\end{align}

The associated rates of heat production per unit mass are related by $dQ_\|/dt=T_\|ds_\|^p/dt$ and $dQ_\perp/dt=T_\perp ds_\perp^p/dt$. We thus get, for the total heat production $Q=Q_\|+Q_\perp$
\begin{equation}
\partial_t(\rho Q)+\nab\cdot\left ( \rho Q \boldsymbol{u}\right )=-\frac{q_\|}{2}(\widehat{\textbf{b}}\cdot\nab)\ln T_\|-q_\perp(\widehat{\textbf{b}}\cdot\nab)\ln T_\perp .
\end{equation}
The global heating is thus given by
\begin{equation}
H=-\int  \left (\frac{q_\|}{2}(\widehat{\textbf{b}}\cdot\nab)\ln T_\|+q_\perp(\widehat{\textbf{b}}\cdot\nab)\ln T_\perp\right ) d^3x, \label{eq:H}
\end{equation}
where $q_\|$ and $q_\perp$ are the heat fluxes obtained from the integration of the model closed at the level of the fourth-rank moments.

We can define a spectral density for the heating rate $H$ (also referred to as  co-spectrum) in the form
\begin{widetext}
\begin{equation}\label{hflux}
H({\boldsymbol k})=-\frac{1}{2}\left ( \frac{1}{2}{\cal F}\{q_\|\}(-{\boldsymbol k}){\cal F}\left \{ (\widehat{\textbf{b}}\cdot\nab)\ln T_\| \right \}({\boldsymbol k})+{\cal F}\{q_\perp\}(-{\boldsymbol k}){\cal F}\left \{ (\widehat{\textbf{b}}\cdot\nab)\ln T_\perp \right \}({\boldsymbol k}) + c.c.\right )
\end{equation}
\end{widetext}
where ${\cal F}$ denotes the Fourier transform.

A few remarks can be made here:\\
1. In all the simulations we have performed, the volume integrated heat production is observed to be positive but its pointwise value can be negative in relatively small regions of space. This contrasts with the (semi-)collisional regime where the heat fluxes obey  Fourier laws of the form $q=-\kappa (\widehat{\textbf{b}}\cdot\nab) T$, making the heat production positive everywhere in space.\\
2. Inserting in Eq. (\ref{eq:H}) the quantities $q_\perp$ and $q_\|$ obtained by the integration of the dynamical equation for the heat fluxes results in taking into account in the heating rate contributions originating from the heat flux present when a quasi-normal closure is implemented (i.e. where the fourth-rank cumulants are taken equal to zero, thus making the Landau damping disappear). In the present simulations, this contribution does not exceed $15\%$ of the total heating rate. In order to only deal with the heat flux originating from the Landau damping, it would be necessary to define a conserved entropy for the quasi-normal closure and evaluate its rate of change due to the introduction of Landau damping. This is left for future work as it is not  straightforward.\\
3. More importantly, this heating rate takes into account the Landau damping on all the waves present in the simulations, including the magnetosonic  waves. At this level, it appears difficult to separate the contributions of the KAW and to evaluate their dissipation by Landau damping. Nevertheless, these magnetosonic waves get dissipated at large scales, thus at small enough scales the estimated heating rate mostly results from Landau damping of KAWs and it becomes possible to compare it to the cascading energy. This particularity is also the reason why Landau damping appears to be acting at all scales in all the results presented above, even in simulations forced at large scales.

\begin{figure}
	\includegraphics[width=\hsize,trim={50 10 20 20},clip]{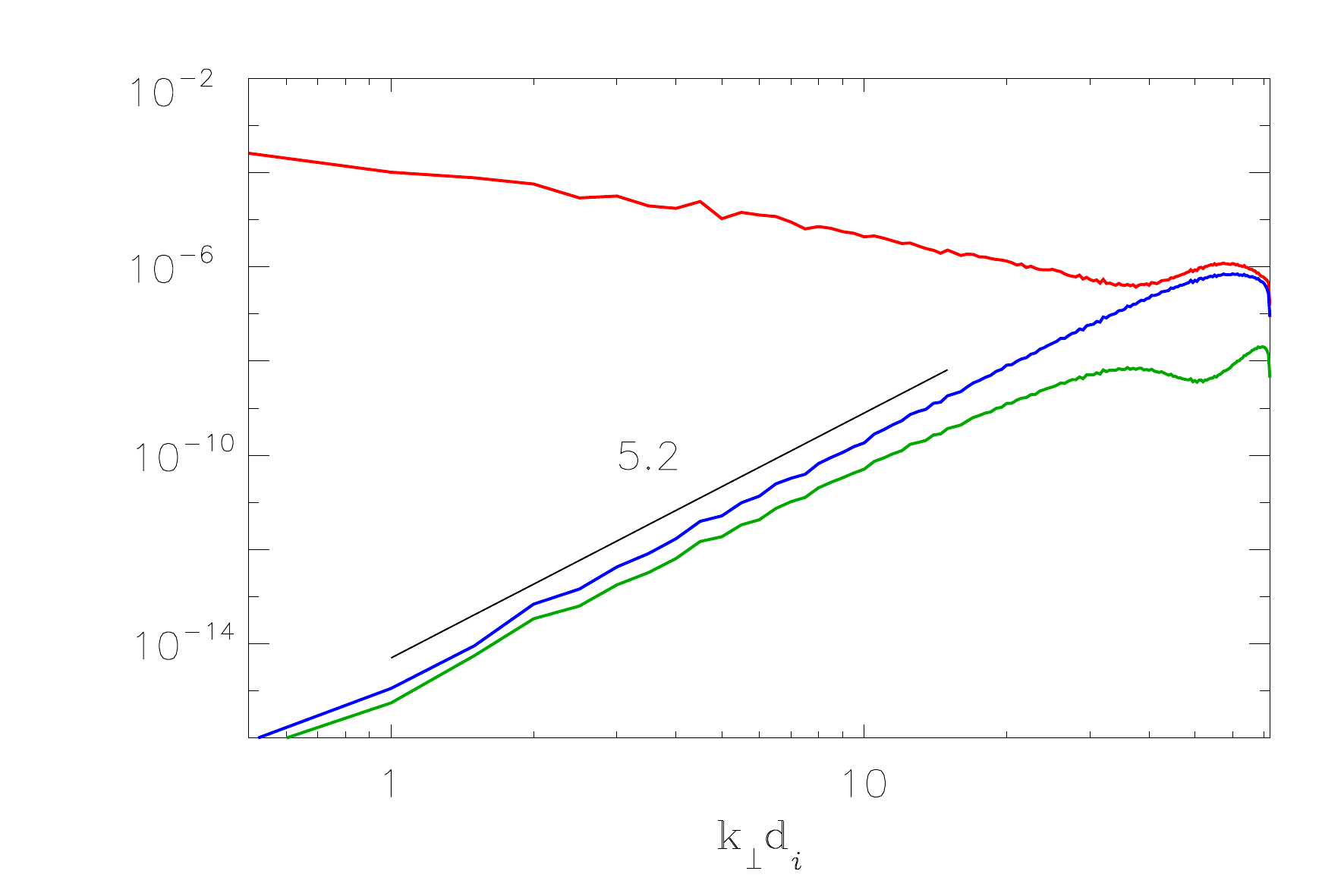}
	\caption{Spectral densities of the heating rate $D_{\rm L}(k_\perp)$ (red) and of the magnetic (blue) and kinetic (green) hyper-dissipation as functions of the transverse wavenumber $k_\perp$ for run LF3. A straight line of slope 5.2 is supplemented, for comparison with the scale-variation of the magnetic hyper-dissipation.}
	\label{OCA_spec_dens}
\end{figure}

The fact that Landau damping is present at all scales in the simulation can be seen by estimating the spectral density of total heating rate at a given wavenumber $k_\perp$, $D_{\rm L}(k_\perp) = \int k_\perp H({\boldsymbol k}) dk_z d\theta$, where $H({\boldsymbol k})$ is the sum of the spectral densities given by equation (\ref{hflux}) for both the ions and the electrons. This spectral density is represented in Fig. \ref{OCA_spec_dens} along with the densities of hyperdissipation and hyperdiffusivity. One clearly sees that the heating rate due to the presence of heat fluxes dominates hyper-dissipation over a broad range of scales due to the dissipation of KAWs and magnetosonic modes, the two becoming comparable only at the smallest scales (note that the magnetic hyper-dissipation dominates at small scales over the kinetic one). 

\subsection{Dissipation due to Landau damping}

The energy ${\mathcal E}_\perp (t)$ of the  (quasi-incompressible) KAWs that cascade towards small scales, and which is the subject of our study, obeys
\begin{equation}
\frac{d}{dt} {\mathcal E}_\perp (t)= {\mathcal I}^C(t)-{\mathcal D}^C_{\rm L}(t)-{\mathcal D}^C_{\rm h}(t),
\end{equation}
where ${\mathcal I}^C$ is the part of the injection rate that contributes to the KAW cascade (the other part is transferred to magnetosonic modes which are dominantly dissipated at large scales), while ${\mathcal D}^C_{\rm L}$ and ${\mathcal D}^C_{\rm h}$ are the parts of the Landau and hyperviscous (and hyperdiffusive)  dissipation that affect the cascading modes. Using cylindrical coordinates and assuming time stationarity, one can write the integrated energy balance at each Fourier mode as (adopting roman scripts for spectral densities): 
\begin{equation}
\epsilon (k_\perp)=\int_0^{k_\perp} \left \{ I^C(k'_\perp) -D^C_{\rm L}(k'_\perp)-D^C_{\rm h}(k'_\perp) \right \} d k'_\perp .
\end{equation}
Considering two wavenumbers $k_{\perp 1}$ and $k_{\perp 2}$ large enough so that the forcing (which is concentrated at large scales) leads to $\int_0^{k_{\perp1}} I^C(k'_\perp) d k'_\perp =\int_0^{k_{\perp2}} I^C(k'_\perp) d k'_\perp ={\mathcal I}^C$, yet small enough for hyperviscous dissipation to be negligible, one obtains:
\begin{equation}
\epsilon(k_{\perp 1}) -\epsilon(k_{\perp 2})=\int_{k_{\perp 1}}^{k_{\perp 2}} D^C_{\rm L}(k'_\perp)d k'_\perp \lesssim \int_{k_{\perp 1}}^{k_{\perp 2}} D_{\rm L}(k'_\perp)d k'_\perp. \label{eq:eps-landau}
\end{equation}
The inequality draws closer to an equality for values of $k_\perp$ large enough so that all magnetosonic modes have been dissipated.

\begin{figure}
\centering
\includegraphics[width=\hsize,trim={70 0 110 60},clip]{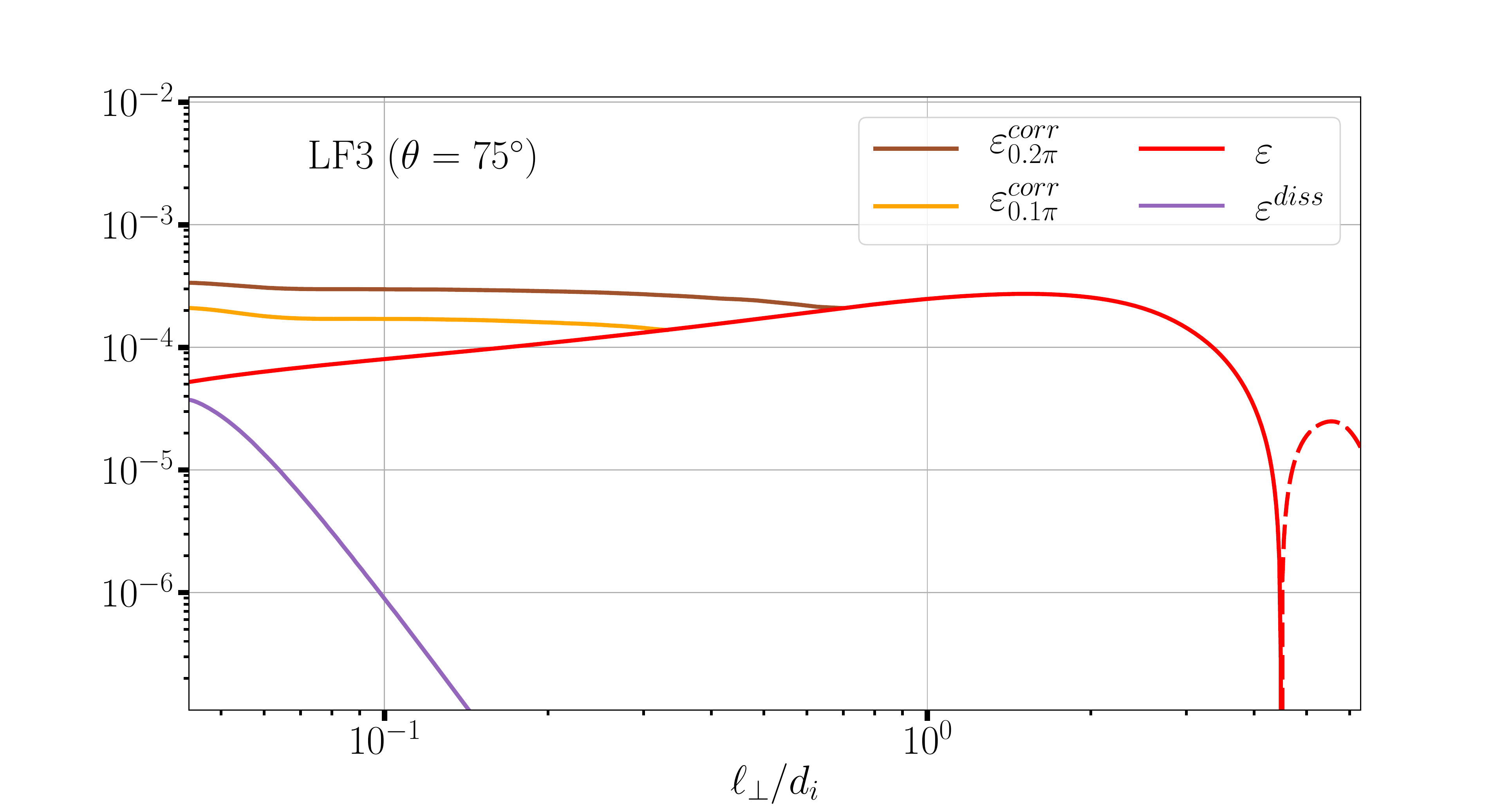}
\caption{Energy cascade rate $\varepsilon$ (red) and transverse hyperdissipation (violet) for run LF3. The orange and brown curves show the same $\varepsilon$ corrected by Landau damping integrated between $\ell_{\perp}$ and a reference scale $\ell_{\perp}=0.1\pi d_i$ and $\ell_{\perp}=0.2\pi d_i$ respectively.}
\label{LFcorr}
\end{figure}

Equation (\ref{eq:eps-landau}) can be used to estimate a correction to the energy cascade rate which would take into account the energy lost due to Landau damping. We do so for run LF3: using this equation we add to the transfer rate the cumulative Landau dissipation between an arbitrary scale (chosen however to be not too large nor too small) and the running (smaller) scale $l_\perp$. Two of these resulting corrected rates $\varepsilon^{corr}$ are shown in Fig. \ref{LFcorr}. They appear to be almost constant, and as such they behave very similarly to the transfer rate of run CGL3 (Fig. \ref{small}). The slight increase of $\varepsilon^{corr}$ towards small scales probably reflects the (weak) contribution of some remaining magnetosonic waves to the calculated Landau damping. This clearly demonstrates that the energy lost along the cascade due to Landau damping is well captured by the decline of the (fluid) cascade rate at the corresponding scales. 

\begin{figure}
	\includegraphics[width=\hsize,trim={35 10 20 20},clip]{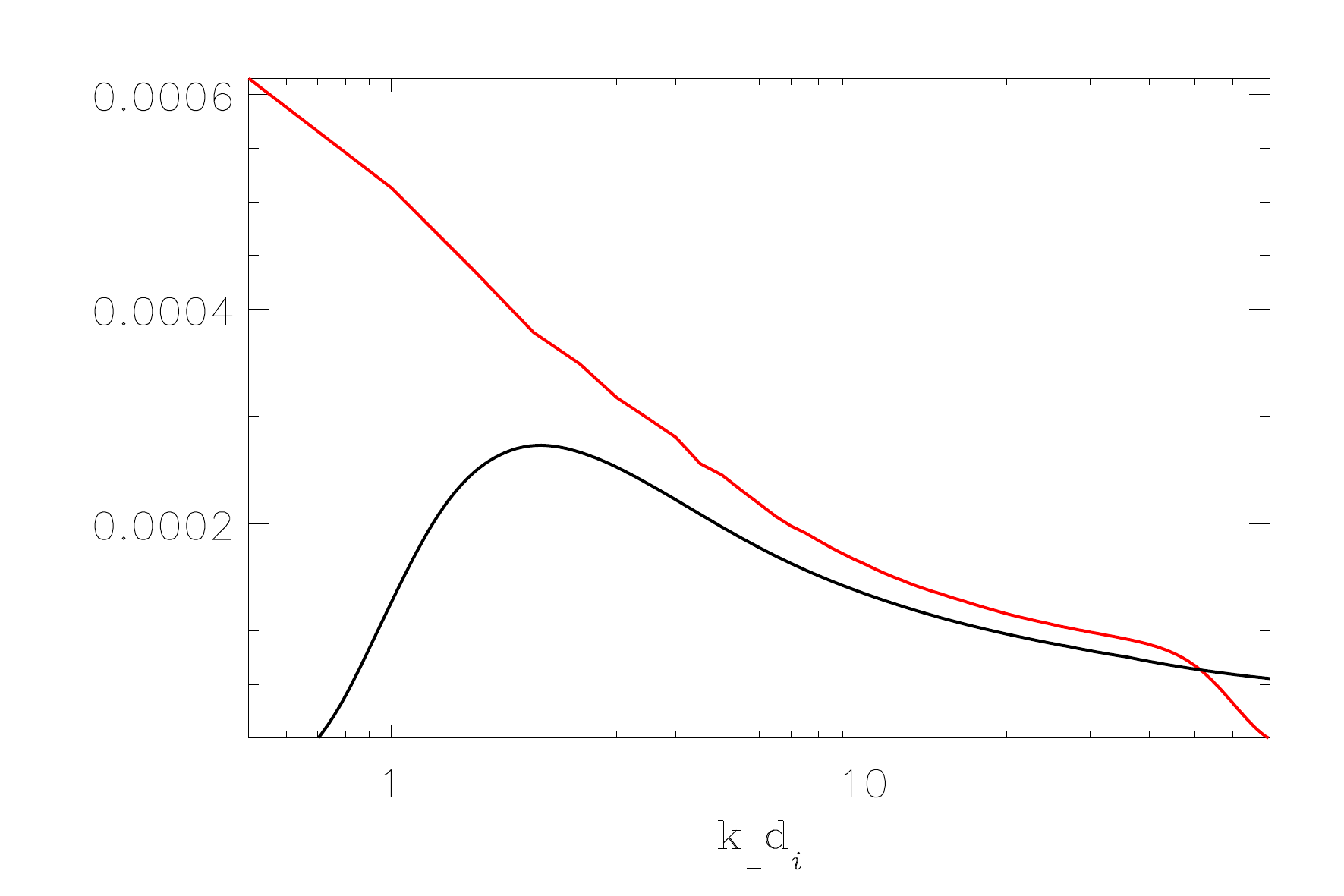}
	\caption{Energy cascade rate $\epsilon(k_\perp)$ (black line) together with Landau and hyper-dissipation (red line) computed with equation (\ref{eq:eps-diss}) for run LF3.}
	\label{LFcorr_spec}
\end{figure}

A complementary estimate of energy dissipation can be done in Fourier space by also taking into account hyperdissipation. Indeed, assuming stationarity, one can also derive that
\begin{widetext}
\begin{equation}
\epsilon (k_{\perp }) ={\mathcal I}^C-{\mathcal D}^C_{\rm L}+ \int_{k_{\perp }}^{\infty}  D^C_{\rm L}(k'_\perp)d k'_\perp-\int_{0}^{k_{\perp }}  D^C_{\rm h}(k'_\perp)d k'_\perp=\int_{k_{\perp }}^{\infty} \left \{ D^{\rm C}_{\rm h}(k'_\perp)+  D^C_{\rm L}(k'_\perp)\right \} d k'_\perp.
\label{eq:eps-diss}
\end{equation}
\end{widetext}
Equation (\ref{eq:eps-diss}) indicates that, as expected, the rate of energy
transfer at the  wavenumber  $k_{\perp}$ identifies with the sum of the rates of Landau and hyperdissipation beyond this wavenumber. One can compare the second right-hand-side term of this equation to the energy cascade rate $\epsilon(k_\perp)$ obtained from the IHMHD exact law, as displayed in Fig. \ref{LFcorr_spec}. The difference between the two curves, which is especially significant at large scales, is due to the fact that the estimation of the dissipation includes the Landau damping of magnetosonic modes, whereas the cascade rate considers only incompressible modes. At smaller scales however, where magnetosonic modes have already been dissipated, the dissipation and cascade rates decreases parallel to each other: this indicates that, at scales not yet affected by hyperdissipation, the decay of $\epsilon(k_\perp)$ in a spectral interval identifies with Landau dissipation within this interval.

Figs. \ref{LFcorr} and \ref{LFcorr_spec} clearly demonstrates that, through the cascade, the energy lost due to Landau damping is well captured by the decline of the (fluid) cascade rate at the corresponding scales. Note that a similar decline of the fluid cascade rate at kinetic scales was reported in 2D hybrid PIC simulations and spacecraft observations in the SW and magnetosheath \citep{hellinger18, Bandyopadhyay20}. Also, \citet{sorriso19} found a correlation between enhancement of a proxy of the local cascade rate and the signatures of wave-particle interactions in MMS data.

\section{Conclusion} \label{conclusion}

In this study, we tackle a fundamental question about the ability of fluid exact laws to reflect the presence of kinetic (Landau) damping. By constructing multi-scale energy cascade and dissipation rates using the HMHD model on a variety of turbulence simulations bearing different intensities of Landau damping, we showed that the presence of Landau damping at small (kinetic) scales is reflected by the steady decline of the energy cascade rate at the same scales, which was found to be comparable to the effective Landau dissipation at those scales. By demonstrating the ability of a fluid exact
law to provide a correct estimate of kinetic dissipation in the sub-ion range of numerical simulations, this work provides a means to evaluate the
amount of energy that is dissipated into particle heating in
spacecraft data: the decline of the cascade rate
allows one to evaluate the kinetic dissipation as a function of scale.
This should help investigating (at least partially) a longstanding
problem in astrophysical plasmas about energy partition
between ions and electrons \citep{Kawazura19}, which are generally heated at different scales. 

The study presented in this paper only makes use of Landau damping. It would be interesting in future works to extend these conclusions to a broad variety of kinetic effects and to test them on more general simulations of the SW, featuring a plasma turbulence driven by other types of waves than slightly perturbed KAWs. It is also important to stress that, even if the oversimplified (yet fully nonlinear) fluid models of turbulence can provide good estimates of the amount of energy that is dissipated into particle heating, they do not specify how this dissipation occurs. The answer to this question and those related to the fate of energy when handed to the plasma particles requires a kinetic treatment.



\section{Acknowledgements} 
This work was granted access to the HPC resources of CINES/IDRIS under the allocation A0060407042. Part of the computations have also been done on the ``Mesocentre SIGAMM" machine, hosted by Observatoire de la C\^ote d'Azur. 
\bibliography{Ref}
\end{document}